\input harvmac
\def\ad{{\dot a}}

\def\bd{{\dot b}}
\def\l{{\lambda}}

\def\d{{\delta}}
\def\e{{\epsilon}}
\def\s{{\sigma}}

\def\half{{1\over 2}}
\def\p{{\partial}}

\def\t{{\theta}}

\def\bar{\overline}

\Title{\vbox{\hbox{IFT-P.048/2004, TCD-Math-04-16, HMI-04-05 }}} {\vbox{ 
\centerline{\bf Higher-Dimensional Twistor Transforms using Pure Spinors}
}}
\bigskip\centerline{Nathan Berkovits$^*$
}
\bigskip
\centerline{\it Instituto de F\'\i sica Te\'orica, Universidade
Estadual Paulista} \centerline{\it Rua Pamplona 145, 01405-900,
S\~ao Paulo, SP, Brasil} \smallskip\centerline{and
}\smallskip\centerline{Sergey A. Cherkis$^{\#}$}
\bigskip
\centerline{\it Institute for Advanced
Study, Princeton NJ 08540, USA}
\centerline{\it Trinity College, Dublin 2, Ireland}
   \vskip .3in
\noindent Hughston has shown that projective pure spinors can be
used to construct massless solutions in higher dimensions,
generalizing the four-dimensional twistor transform of Penrose. In
any even (Euclidean) dimension $d=2n$, projective pure spinors
parameterize the coset space $SO(2n)/U(n),$ which is the space of
all complex structures on ${\bf R}^{2n}$. For $d=4$ and $d=6$,
these spaces are ${\bf CP}^1$ and ${\bf CP}^3 $ and the
appropriate twistor transforms can easily be constructed. In this
paper, we show how to construct the twistor transform for $d>6$
when the pure spinor satisfies nonlinear constraints, and present
explicit formulas for solutions of the massless field equations.

   \vskip 1cm

* {e-mail: nberkovi@ift.unesp.br},   \# e-mail: cherkis@ias.edu

\Date{September 2004}

\newsec{Introduction}

As defined by Cartan \ref\Cartan{E.~Cartan, The Theory of Spinors (Dover,
New York,
1981).} and Chevalley \ref\Chevalley{C.~Chevalley, The Algebraic Theory of
Spinors
(Columbia University Press, New York, 1954).}, pure spinors in even
dimension $d=2n$ are
complex spinors $\l^a$ which satisfy the constraint $\l^a (\s^{\mu_1 ..
\mu_j})_{ab}
\lambda^b =0$ for $0\leq j< n$, where $\s^{\mu_1 ...\mu_j}$ is the
antisymmetrized
product of $j$ Pauli matrices. So $\l^a \l^b$ can be written as \eqn\pures
{\l^a \l^b =
{{1}\over{n! ~2^n}}\s_{\mu_1 ...\mu_n}^{ab} ~(\l^c \s^{\mu_1 ..\mu_n}_{cd}
\l^d)} where
$\l\s^{\mu_1 ... \mu_n} \l$ defines an $n$-dimensional complex plane,
and thus complex coordinates on ${\bf R}^{2n}$. In Euclidean space, this
$n$-dimensional complex plane is preserved up to a phase by a $U(n)$
subgroup of $SO(2n)$
rotations. So projective pure spinors in $d=2n$ Euclidean dimensions
parameterize the
coset space $SO(2n)/U(n)$.\foot{In
Minkowski
space, the $n$-dimensional complex plane is preserved by a $U(n-1)$ subgroup
of $SO(d-2)$
rotations and is also preserved by $(2n-1)$ light-like boosts. So projective
pure spinors
in Minkowski space contain the same number of variables as in Euclidean
space, but the
coset space is modified to $SO(2n-1,1)/U(n-1)\times {\bf R}^{2n-1}$.}
(For a more detailed account of this correspondence,
we refer the
reader to the Appendix.)

In four dimensions, this is the coset space $SO(4)/U(2)={\bf
CP}^1$ which is parameterized by a projective chiral spinor $\l^a$
for $a=1$ to 2. As is well-known, the twistor formalism of Penrose
makes use of this $d=4$ projective pure spinor to construct
solutions to $d=4$ massless equations of motion through a twistor
transform formula \ref\twistorfoura {R.~Penrose, ``The nonlinear
graviton,''  General Relativity and Gravitation  7  (1976), no. 2,
171--176.} \ref\twistorfourb{ N.~J.~Hitchin, ``Linear field
equations on self-dual spaces,'' Proc. Roy. Soc. London Ser. {\bf
A 370} (1980), no. 1741, 173--191.}. In six dimensions, the coset
$SO(6)/U(3) = {\bf CP}^3$ is parameterized by a projective chiral
spinor $\l^a$ for $a=1$ to 4. Although it is less well-known than
its four-dimensional counterpart, this projective pure spinor in
six dimensions can similarly be used to construct solutions to the
$d=6$ massless equations of motion through a twistor transform
formula, as demonstrated by Hughston \ref\twistorsix{ R.~Penrose
and W.~Rindler, ``Spinors And Space-Time. Vol. 2: Spinor And
Twistor Methods In Space-Time Geometry,'' Cambridge University
Press, 1986 (See the Appendix).}
\ref\twistorsixorig{L.~P.~Hughston,``Applications of SO(8) Spinors,'' 
pp. 253-287 in Gravitation and
Geometry: a volume in honour of Ivor Robinson (eds. W. Rindler and
A. Trautman), Bibliopolis, Naples (1987). }.

Hughston \ref\hugh {L.P. Hughston, ``The Wave Equation in Even
Dimensions,'' in Further Advances in Twistor Theory, vol. 1,
Research Notes in Mathematics 231, Longman, pp. 26-27, 1990\semi
L.P. Hughston, ``A Remarkable Connection between the Wave Equation
and Pure Spinors in Higher Dimensions,'' in Further Advances in
Twistor Theory, vol. 1, Research Notes in Mathematics 231,
Longman, pp. 37-39, 1990\semi L.P. Hughston and L.J. Mason, ``A
Generalized Kerr-Robinson Theorem,'' Classical and Quantum Gravity
5 (1988) 275. } has also argued that twistor transform
constructions of solutions to massless equations of motion can be
generalized using projective pure spinors in any even dimension.
Above six dimensions, the construction becomes non-trivial since
pure spinors for $d\geq 8$ satisfy nonlinear constraints. For
example, in eight dimensions, the coset $SO(8)/U(4)$ is
parameterized by a projective chiral spinor $\l^a$ for $a=1$ to 8
satisfying the additional constraint $\l^a \l^a=0$. And in ten
dimensions, the coset $SO(10)/U(5)$ is parameterized by a
projective chiral spinor $\l^a$ for $a=1$ to 16 satisfying the
constraint $\l^a \s^\mu_{ab}\l^b=0$ where $\s^\mu_{ab}$ are the
$d=10$ Pauli matrices. So generalization of the twistor transform
construction to higher dimensions requires new techniques for
integration over these coset spaces. In this paper, these
integration techniques will be developed and explicit twistor
transform formulas will be constructed using pure spinors in any
even dimension. In particular, the covariant measure is written
explicitly in terms of pure spinors. As an example, these formulas
provide higher-dimensional scalar Green's functions.


Here we limit ourselves to the simple question of solving linearized
massless equations
of motion in flat spacetime. However, the fact that projective pure spinors
provide an
elegant higher-dimensional generalization of this twistor construction
suggests that pure
spinors may also be useful for generalizing other applications of four-
dimensional
twistors to higher dimensions. Solutions of nonlinear problems, as well as
linear
problems in non-flat background, are of special interest. For example, four-
dimensional
twistors have been useful for constructing solutions of self-dual Yang-Mills
\ref\ADHM{M.~F.~Atiyah, N.~J.~Hitchin, V.~G.~Drinfeld and Y.~I.~Manin,
``Construction of
instantons,'' Phys.\ Lett.\ A {\bf 65}, 185 (1978).} and self-dual gravity
equations
\ref\twistor1
\ref\NH{N.~J.~Hitchin, ``Twistor construction of Einstein metrics,'' Global
Riemannian geometry (Durham, 1983), 115--125, Ellis Horwood Ser. Math.
Appl., Horwood,
Chichester, 1984. }
\ref\CH{S.~A.~Cherkis and N.~J.~Hitchin, ``Gravitational instantons
of type D(k),'' arXiv:hep-th/0310084.} ,
and for constructing Green's functions on
multi-Taub-NUT spaces
\ref\Atiyah{ M.F. Atiyah ``Green's functions for self-dual
four-manifolds,'' Math. Analysis and Applications, Part A Advances in
Mathematics
Supplementary Studies, Vol 7A (1981) 129-158. }.
It might be possible that pure spinors will be
useful for generalizing these nonlinear constructions to higher dimensions.
In
particular, we use pure spinors in this paper to construct self-dual abelian
potentials
in higher dimensions and one might hope that geometric insight from this
construction
will lead to the proper formalism for a nonabelian generalization.

Note that ten-dimensional pure spinors have recently been used for
covariantly
quantizing the superstring \ref\tenpure{N. Berkovits,
``Super-Poincar\'e covariant quantization of the
superstring,'' JHEP 04 (2000) 018, hep-th/0001035.}\ref\loop{
N. Berkovits, ``Multiloop amplitudes and vanishing theorems using
the pure spinor formalism for the superstring,'' hep-th/0406055.}
and many of the techniques described here
are generalizations of techniques
developed for quantization of the ten-dimensional superstring.
So it would not be surprising to find that pure spinors in ten dimensions
are useful for constructing supersymmetric
solutions to $d=10$ super-Yang-Mills and
supergravity equations
\ref\howet{P. Howe, ``Pure spinors, function superspaces and supergravity
theories in ten-dimensions and eleven-dimensions,'' Phys. Lett. B273
(1991) 90\semi P. Howe, ``Pure spinors lines in superspace and
ten-dimensional
supersymmetric theories,'' Phys. Lett. B258 (1991) 141.},
which are the low-energy equations of the
superstring.
However, it is not clear how the higher-dimensional twistors described
here can be generalized to higher-dimensional supertwistors.

There have been numerous approaches to generalizing the
twistor formalism to higher
dimensions, most of which
differ from each other and from our approach.
For example, Ward presents classes of various nonlinear
equations for a nonabelian gauge field
in \ref\Ward{ R.~S.~Ward,
``Completely
solvable gauge-field equations in dimension greater than four,''  Nucl.\
Phys.\ {\bf B
236} (1984),  no. 2, 381--396. }
that can be solved using higher-dimensional twistors. Also, twistor-like
transforms in higher dimensions have appeared in studies of the
superparticle and superstring (e.g.
\ref\supers{
E. Witten, ``Twistor-like transform in ten dimensions,'' Nucl. Phys. B266
(1986) 245\semi
A. Bengtsson, I. Bengtsson, M. Cederwall and N. Linden,
``Particles, superparticles and twistors,'' Phys. Rev. D36 (1987) 1766\semi
E. Sokatchev, ``Harmonic superparticle,'' Class. Quant. Grav. 4 (1987)
237\semi
D. Sorokin, V. Tkach and D. Volkov, ``Superparticles,
twistors and Siegel symmetry,'' Mod. Phys. Lett. A4 (1989) 901\semi
N. Berkovits, ``A supertwistor description of the massless superparticle
in ten-dimensional superspace,'' Nucl. Phys. B350 (1991) 193\semi
E. Bergshoeff, P. Howe, C. Pope, E. Sezgin
and E. Sokatchev, ``Ten-dimensional supergravity from lightlike
integrability in loop superspace,'' Nucl. Phys. B354 (1991) 113\semi
A. Galperin, P. Howe and P. Townsend, ``Twistor transform for superfields,''
Nucl. Phys. B402 (1993) 531.}\howet).
The resemblance of pure spinors and twistors
has been noted by many people (e.g. \ref\budinich{P. Budinich and
A. Trautman, The Spinorial Chessboard (Trieste Notes in Physics,
Springer-Verlag, Berlin, 1988)
\semi
P. Budinich, ``From the geometry of pure spinors with their
division algebras to fermion's physics,'' Found. Phys. 32 (2002) 1347,
hep-th/0107158.}\ref\furlan{ P. Furlan and R.
Raczka, ``Intrinsic nonlinear spinor wave equations
associated with nonlinear spinor representations,''
Journal of Mathematical Physics 27 (1986) 1883\semi  P.Furlan and R.Raczka,
``Nonlinear  Spinor Representations,''
Journal of Mathematical Physics 26 (1985) 3021.}
\ref\harnad{J.~P.~Harnad and S.~Shnider,
``Isotropic Geometry, Twistors And Supertwistors. 1. The Generalized Klein
   Correspondence And Spinor Flags,''
J.\ Math.\ Phys.\  {\bf 33}, 3197 (1992).}) and, 
aside from the Hughston
papers in \hugh,
the references we are aware of which
come closest to the explicit approach presented here are
\ref\OBR{M.~R.~O'Brian and
J.~H.~Rawnsley, ``Twistor spaces,''
Ann. Global. Anal. Geom. {\bf 3} (1985) 29-58.} where the
properties of
twistor space are studied, as well as \ref\Murray{M.~K.~Murray, ``A
Penrose transform
for the twistor space of an even-dimensional conformally flat Riemannian
manifold,'' Ann.
Global Anal. Geom.  {\bf 4}  (1986),  no. 1, 71--88.} and \ref\Inoue
{Y.~Inoue, ``Twistor
spaces of even-dimensional Riemannian manifolds,'' J. Math. Kyoto Univ. 32
(1992), no. 1,
101--134.} where the Penrose transform is constructed and proven to be one-
to-one.

In section 2 of this paper, we review the twistor construction of
massless solutions using four and six-dimensional pure spinors. In
section 3,
we show how this twistor construction extends to pure spinors in eight and
ten
dimensions. And in section 4, we
generalize this twistor construction to pure spinors in arbitrary even
dimensions.

\newsec{Pure Spinors in Four and Six Dimensions}

In four and six dimensions, projective pure spinors parameterize the coset
spaces
$SO(4)/U(2) = {\bf CP}^1$ and $SO(6)/U(3)={\bf CP}^3$, and are therefore
described by
complex projective two-component and four-component spinors which have been
called
twistors. As will be reviewed here, these twistors have been used for
constructing
solutions to massless equations of motion in four \twistorfoura\twistorfourb
and six
dimensions \twistorsix.

\subsec{Four dimensions}

As is well-known, Penrose has used complex projective two-component spinors
to construct
twistor solutions to massless equations of motion in four dimensions
\twistorfoura\twistorfourb. To describe this method in a manner which will
generalize to
higher dimensions, consider the massless Klein-Gordon equation $\p^\mu
\p_\mu \Phi(x)=0$
for a scalar field $\Phi(x)$ where $\mu=1$ to 4. It is useful to combine
$x^\mu$ into a
pair of complex coordinates, $z_1 = x_1 + i x_2$ and $z_2 = x_3 +i x_4$, so
that the
Klein-Gordon equation (in Euclidean space\foot {As usual, one can Wick
rotate to
Minkowski space by replacing $x_4$ with $ix_4$ so that $z_2 = x_3 + x_4$ and
$\bar z_2 =
x_3-x_4$ are independent real variables.}) is $\p_{z_j}\bar\p_{\bar z_ j}
\Phi(z,\bar z)
=0$. Then if one defines \eqn\fourone{ w_1 = z_1 + u\bar z_2,\quad w_2 =
z_2 - u\bar z_1}
where $u$ is a complex variable, any holomorphic function $f(w_1,w_2)$ will
satisfy
\eqn\satisfy{(\p_{z_1}\bar\p_{\bar z_1} + \p_{z_2} \bar\p_{\bar z_2}) f
(w_1,w_2)= \left(
{\p\over{\p w_1}} (-u {\p\over{\p w_2}}) + {\p\over{\p w_2}} (u{\p\over{\p
w_1}}) \right)
f(w_1,w_2)=0.} So the massless Klein-Gordon equation has the solution
\eqn\foursol{
\Phi(z,\bar z) = \oint du f(u, w_1,w_2)|_{w_1 = z_1 + u\bar z_2,w_2 = z_2 -
u\bar z_1}}
where $\oint du$ is a contour integral around any region in the complex
plane.

This construction of massless $d=4$ solutions can be made manifestly
Lorentz covariant by introducing a bosonic
projective spinor $\l^a$ for $a=1$ to 2
and defining
\eqn\fourtwo{ w_\ad = \s^\mu_{a\ad} x_\mu \l^a}
where $\s^\mu_{a\ad}$ are the usual $d=4$ Pauli matrices.
Under $d=4$ conformal transformations, $(\l^a, w_\ad)$ transforms
linearly as an $SO(4,2)$ spinor.

When $\l^a = (1,u)$, the relation of \fourtwo\ reduces to \fourone\
and solution \foursol\ can be written covariantly as the twistor
transform formula
\eqn\foursoltwo{ \Phi(x) = \oint d\l^a ~\l_a  F(\l,w)|_{w=x \l}}
where $F(h\l^a,h w_\ad) = h^{-2} F(\l^a,w_\ad)$ so that the contour integral
over the projective spinor is well-defined.
For example, choosing
\eqn\fourextwo{ F(\l,w) =  {{\e_{\ad\bd} A_1^\ad A_2^\bd}\over
{(A_1^{\dot c} w_{\dot c})(A_2^{\dot d} w_{\dot d})}}  }
generates the $d=4$ Green's function
$\Phi(x) = (x^\mu x_\mu)^{-1}.$

One can similarly construct massless $d=4$ solutions to higher-spin
equations by considering functions $F(\l^a, w_\ad)$ satisfying the
condition
$F(h\l^a,h w_\ad) = h^{-N-2} F(\l^a,w_\ad)$.
If $N$ is positive, one uses the twistor transform formula
\eqn\foursolthree{ \Phi^{(a_1 ... a_{N})}(x)
= \oint d\l^b ~\l_b \l^{a_1} ...\l^{a_{N}} F(\l,w)|_{w=x\l}.}
And if $N$ is negative, one uses the formula
\eqn\foursolfour{ \Phi^{(\ad_1 ... \ad_{-N})}(x)
= \oint d\l^a ~\l_a  \left({\p\over\p w_{\ad_1}} ...
{\p\over\p w_{\ad_{-N}}} F(\l,w)\right)|_{w=x\l}.}
Since ${\p\over{\p x^\mu}} F(\l,w) = (\l\s_\mu)_\ad {\p\over{\p w_\ad}}F
(\l,w)$,
one can use $\s^\mu_{a\ad} \s_{\mu~b {\dot b}} =2\e_{ab}\e_{\ad{\dot b}}$ to
show that
$\s^\mu_{b\dot b}{\p\over{\p x^\mu}}
\Phi^{(b a_2 ... a_{N})}(x)=0$
and
$\s^\mu_{b\dot b}{\p\over{\p x^\mu}}
\Phi^{(\dot b \ad_2 ... \ad_{-N})}(x)=0$.
So \foursolthree\ and \foursolfour\ describe solutions for massless
particles of spin $|N|/2$ and helicity $N/2$.

\subsec{Six dimensions}

Although less familiar than the two-component
twistor formulas in four dimensions,
projective four-component
complex spinors have been used to
construct twistor solutions to massless equations of motion
in six dimensions \twistorsix.
For example, consider the Klein-Gordon massless
equation $\p^\mu \p_\mu \Phi(x)=0$ for a scalar field $\Phi(x)$ where
$\mu=1$ to 6. As before,
combine $x^\mu$ into a triplet of complex coordinates,
$z_1 = x_1 + i x_2$, $z_2 = x_3 +i x_4$ and
$z_3 = x_5 +i x_6$, so that the Klein-Gordon equation (in Euclidean space)
is $\p_{z_j}\bar\p_{\bar z_j} \Phi(z,\bar z) =0$.
Then if one defines
\eqn\sixone{ v_j = z_j + u_{jk}\bar z_k}
where $u_{jk} = -u_{kj}$ are three independent complex variables,
any holomorphic function $f(v_1,v_2, v_3)$
will satisfy
\eqn\sixsatisfy{\p_{z_j}\bar\p_{\bar z_j} f(v_1,v_2, v_3)=
u_{jk} {\p\over{\p v_j}} {\p\over{\p v_k}}
f(v_1,v_2, v_3)=0}
because of the antisymmetry of $u_{jk}$. So
the massless Klein-Gordon equation has the solution
\eqn\sixsol{ \Phi(z,\bar z) = (\oint du)^3 f(u_{jk}, v_l)|_{v_j
= z_j + u_{jk}\bar z_k}}
where the three contour integrals for the $u_{jk}$ variables are chosen
arbitrarily.

This construction of massless $d=6$ solutions can be made manifestly
Lorentz covariant by introducing a projective spinor $\l^a$ for $a=1$ to 4
and defining
\eqn\sixtwo{ w_a = \s^\mu_{ab} x_\mu \l^b}
where $\s^\mu_{ab} = -\s^\mu_{ba}$ are the $d=6$ Pauli matrices.
Under $d=6$ conformal transformations, $(\l^a, w_\bd)$ transforms
linearly as an $SO(6,2)$ spinor.

When $\l^a = (1,u_{23}, u_{31}, u_{12})$, one can check that
with a suitable choice for the Pauli matrices,
\eqn\sixsuit{ w_a = (\half \e^{jkl} u_{jk} z_l,~ v_1,~ v_2,~ v_3)}
where $v_j$ is defined in \sixone. Note that \sixsuit\ satisfies
$\l^a w_a=0$, as implied by \sixtwo.
Furthermore, the massless solution \sixsol\ can be covariantly
written as the twistor transform formula
\eqn\sixsoltwo{ \Phi(x) = \oint \e_{abcd} d\l^a \wedge d\l^b \wedge
d\l^c  ~\l^d
F(\l, w)|_{w=x \l}}
where $F(h\l^a,h w_b) = h^{-4} F(\l^a,w_b)$ so that the integral
over the projective spinor is well-defined.
For example, choosing
\eqn\sixextwo{ F(\l,w) =
{\e_{abcd} A_1^a A_2^b A_3^c A_4^d\over
\prod_{r=1}^4 ( A_r^e w_e)}}
generates the $d=6$ Green's function
$\Phi(x) = (x^\mu x_\mu)^{-2}$.

One can similarly construct massless $d=6$ solutions to higher-spin
equations by considering functions $F(\l^a, w_b)$ satisfying the
condition
$F(h\l^a,h w_b) = h^{-N-4} F(\l^a,w_b)$.
When $N$ is positive,
one uses the twistor transform formula
\eqn\sixsolthree{ \Phi^{(a_1 ... a_{N})}(x)
= \oint \e_{bcde}d\l^b\wedge d\l^c \wedge d\l^d ~\l^e  \l^{a_1} ...
\l^{a_{N}} F(\l,w)|_{w=x\l}.}
And when $N$ is negative,
one uses the formula
\eqn\sixsolfour{ \Phi^{(a_1 ... a_{-N})}(x)
= \oint \e_{bcde}d\l^b\wedge d\l^c \wedge d\l^d ~\l^e  \left({\p\over\p
w_{a_1}} ...
{\p\over\p w_{a_{-N}}} F(\l,w)\right)|_{w=x\l}.}
Since ${\p\over{\p x^\mu}} F(\l,w) = (\l\s_\mu)_a {\p\over{\p w_a}}F(\l,w)$,
one can use $\s^\mu_{ab} \s_{\mu~cd} =2\e_{abcd}$ to
show that $\s^\mu_{b c }{\p\over{\p x^\mu}}
\Phi^{(c a_2 ... a_N)}(x)=0$ either when $N$ is positive or negative. So
the solutions of \sixsolthree\ and \sixsolfour\
describe a massless spin $\half$ field when $N=\pm 1$,
a self-dual three-form field-strength when $N=\pm 2$, etc.

\newsec{Pure Spinors in Eight and Ten Dimensions}

Using the methods of the previous section, it is easy to generalize
the non-covariant construction of \foursol\ and \sixsol\ to arbitrary
even dimension.
To solve the massless Klein-Gordon
equation $\p^\mu \p_\mu \Phi(x)=0$ for a scalar field $\Phi(x)$ where
$\mu=1$ to $2n$, first
combine $x^\mu$ into  $n$ complex coordinates,
$z_j = x_{2j-1} + i x_{2j}$ for $j=1$ to $n$, so that
the Klein-Gordon equation in Euclidean space
is $\p_{z_j}\p_{\bar z_ j} \Phi(z,\bar z) =0$.
Defining
\eqn\genone{ v_j = z_j + u_{jk}\bar z_k}
where $u_{jk}= -u_{kj}$ are $n(n-1)/2$ independent complex variables,
one finds that any holomorphic function $f(v_j,u_{jk})$ satisfies
\eqn\gensatisfy{\p_{z_j}\p_{\bar z_j} f(v,u)=
u_{jk}{\p\over{\p v_j}}{\p\over{\p v_k}} f(v, u) =0.}
So
the massless Klein-Gordon equation has the solution
\eqn\gensol{ \Phi(z,\bar z) = (\oint du)^{n(n-1)/2} f(v,u)|_{v_j
= z_j + u_{jk}\bar z_k}}
where the $n(n-1)/2$ contour integrals for $u_{jk}$
are chosen arbitrarily.

To express this solution in a Lorentz-covariant manner using pure spinors,
it will be necessary to know how to
integrate the pure spinors over the coset space
$SO(2n)/U(n)$. When $n=5$, an integration method for pure spinors was
developed in \loop\ for quantization of the ten-dimensional superstring.
As will be shown here, this integration method is easily generalized
for arbitrary $n$, which will allow the massless solution of \gensol\
to be expressed in a Lorentz-covariant manner. Before describing this
twistor transform
construction for arbitrary even dimension, it will be convenient
to first describe the twistor transform construction for $d=8$ and $d=10$.

\subsec{Eight dimensions}

In eight dimensions, a pure spinor is described by a chiral spinor
$\l^a$ for $a=1$ to 8 which satisfies the additional constraint $\l^a \l^a=0
$.
To covariantize $v_j$ and $u_{jk}$ of \genone\ for $j=1$ to 4,
define the antichiral spinor
\eqn\eighttwo{ w_\ad = \s^\mu_{a\ad} x_\mu \l^a}
where $\s^\mu_{a\ad}$ are the $d=8$ Pauli matrices.
Note that $w_\ad$ is an antichiral
pure spinor and under $d=8$ conformal transformations,
$(\l^a, w_\ad)$ transforms linearly as an $SO(8,2)$ spinor.

When $\l^a = (1,~u_{jk},~ -{1\over 8}\e^{jklm} u_{jk} u_{lm})$,
one can choose a representation of the $d=8$ Pauli matrices
such that
\eqn\eightchoose{ w_\ad = (v_j, ~\half\e^{jklm} v_k u_{lm}).}
Note that \eightchoose\
satisfies $\s^\mu_{a\ad} \l^a w^\ad=0$, as implied by \eighttwo.
To covariantize the massless solution of \gensol,
one needs to define a suitable integration measure for integrating
$\l^a$ over the coset space $SO(8)/U(4)$.

To define such an integration measure, note
that
\eqn\eightmeasure{ [d\l]_{d=8}\equiv
(C_b \l^b)^{-1} \e_{a_1 ... a_8} d\l^{a_1} \wedge ... \wedge d\l^{a_6}
\l^{a_7} C^{a_8}}
is independent of the choice of $C^b$ and is therefore Lorentz-invariant.
To show independence of $C^b$, use $\l^a \l^a =0$ and $ \l^a d\l^a=0$ to
show that \eightmeasure\ is invariant under
the transformation
\eqn\eighttransf{\d C_a = f\l_a + g C_a + \e_{a b c_1 ... c_6} \l^b
h^{c_1 ... c_6} }
where
$f$, $g$ and $h^{c_1 ... c_6}$ are arbitrary parameters.
Since \eighttransf\ can be used to change $C_a$ in an arbitrary manner,
\eightmeasure\ is independent of $C_a$.

Using the measure factor of \eightmeasure,
the solution of \gensol\ can be written
in Lorentz-covariant form as the twistor transform formula
\eqn\eightsoltwo{ \Phi(x) = \oint [d\l]_{d=8}  F(\l,w)|_{w=x \l}}
where $F(h\l^a,h w_\ad) = h^{-6} F(\l^a,w_\ad)$ so that the integral
over the projective spinor is well-defined.
For example, choosing
\eqn\eightextwo{ F(\l,w) =
{\e_{\bd_1 ... \bd_8} A_1^{\bd_1} ... A_7^{\bd_7}
w^{\bd_8}\over\prod_{j=1}^7
(A_j^\ad w_\ad)}}
generates the $d=8$ Green's function
$\Phi(x) = (x^\mu x_\mu)^{-3}.$

One can similarly construct massless $d=8$ solutions to higher-spin
equations by using the twistor transform formula
\eqn\eightsolthree{ \Phi^{(a_1 ... a_{N})}(x)
= \oint [d\l]_{d=8} \l^{a_1} ...\l^{a_{N}} F(\l,w)|_{w=x\l}}
where $F(\l^a, w_\ad)$ satisfies the
condition
$F(h\l^a,h w_\ad) = h^{-N-6} F(\l^a,w_\ad)$ for $N$ positive.
Since
$${\p\over{\p x^\mu}} F(\l,w) = (\l\s_\mu)_\ad {\p\over{\p w_\ad}}F(\l,w),$$
$\s^\mu_{a\ad} \s_{\mu~b {\dot b}}\l^a\l^b=0$
implies that
$\s^\mu_{b\dot b}{\p\over{\p x^\mu}}
\Phi^{(b a_2 ... a_{N})}(x)=0$.
So \eightsolthree\
describes a massless spin $\half$ field when $N=1$,
a self-dual four-form field-strength when $N=2$, etc.
But unlike the $d=4$ and $d=6$ cases,
one cannot construct massless solutions when $N$ is negative since
$\s^\mu_{a\ad} \s_{\mu~b {\dot b}}{\p\over{\p w_\ad}}
{\p\over{\p w_\bd}}$ does not necessarily vanish.

\subsec{Ten dimensions}

In ten dimensions, a pure spinor is described by a chiral spinor
$\l^a$ for $a=1$ to 16 which satisfies the additional constraint $\l^a
\s^\mu_{ab} \l^a=0$
where $\s^\mu_{ab}= \s^\mu_{ba}$ are the $d=10$ Pauli matrices.
To covariantize $v_j$ and $u_{jk}$
of \genone\ for $j=1$ to 5, define the antichiral spinor
\eqn\tentwo{ w_a = \s^\mu_{ab} x_\mu \l^b.}
Note that $w_a$ is an antichiral
pure spinor and under $d=10$ conformal transformations,
$(\l^a, w_a)$ transforms linearly as an $SO(10,2)$ spinor.

When $\l^a = (1,~u_{jk}, ~ -{1\over 8}\e^{jklmn} u_{jk} u_{lm})$,
one can choose a representation of the $d=10$ Pauli matrices
such that
\eqn\tenchoose{ w_a = (v_j, ~\half v_{[k} u_{lm]},
~{1\over 8}\e^{jklmn} v_j u_{kl} u_{mn}),}
which satisfies $\l^a w_a = \l^a (\s^{\mu\nu})_a{}^b w_b=0$,
as implied by \tentwo.
To covariantize the massless solution of \gensol,
one needs to define a suitable integration measure for integrating
$\l^a$ over the coset space $SO(10)/U(5)$.

Such a measure was defined in \loop\ as
\eqn\tenmeasure{ [d\l]_{d=10}\equiv
(C_b \l^b)^{-3} \e_{a_1 ... a_{16}}
d\l^{a_1} \wedge ... \wedge d\l^{a_{10}}
\l^{a_{11}} (C\s^\mu)^{a_{12}} (C\s^\nu)^{a_{13}}(C\s^\rho)^{a_{14}}
(\s_{\mu\nu\rho})^{a_{15} a_{16}},}
where $[d\l]_{10}$
is independent of the choice of $C_b$ and is therefore Lorentz-invariant.
A simple way to show independence of the measure of $C_b$ is by using
invariance under the $U(1)\times SU(5)$ subgroup
which preserves the pure spinor $\l^a$ up to a phase.
Under $U(1)\times SU(5)$,
the sixteen components of an $SO(10)$
chiral spinor transform as
$(1_{5/2}, 10_{1/2}, \bar 5_{ -3/2})$ representations and the
sixteen components of an $SO(10)$
antichiral spinor transform as
$(5_{3/2},  \bar{10}_{-1/2}, 1_{-5/2})$ representations, where the
subscript denotes the $U(1)$ charge.
So by our choice of the $U(1)\times SU(5)$ subgroup, $\l^a$ transforms as an
$SU(5)$ singlet with $U(1)$ charge $5/2$.
Furthermore,
since $\l^a \s^\mu_{ab} d\l^b=0$,
$d\l^a$ carries either $U(1)$ charge $5/2$ or $1/2$. Therefore,
$ d\l^{[a_1} \wedge ... \wedge d\l^{a_{10}} \l^{a_{11}]}$ carries
$U(1)$ charge $15/2$, which implies by $U(1)$ conservation
that only the component of $C_b$ in the $1_{-5/2}$ representation
contributes to \tenmeasure. Finally,
it is easy to see that
\tenmeasure\ is invariant under scale transformations of this $1_{-5/2}$
component because there are an equal number of $C_b$'s in the numerator
and denominator of \tenmeasure.

Using the measure factor of \tenmeasure,
the solution of \gensol\ can be written
in Lorentz-covariant form as the twistor transform formula
\eqn\tensoltwo{ \Phi(x) = \oint [d\l]_{d=10}  F(\l,w)|_{w=x \l}}
where $F(h\l^a,h w_b) = h^{-8} F(\l^a,w_b)$ so that the integral
over the projective spinor is well-defined.
For example, choosing
\eqn\tenextwo{ F(\l,w) =
{\e_{b_1 ... b_{16}} A_1^{b_1} ... A_{11}^{b_{11}}
(\s^\mu w)^{b_{12 }} (\s^\nu w)^{b_{13}}(\s^\rho w)^{b_{14}}
(\s_{\mu\nu\rho})^{b_{15} b_{16}}\over
\prod_{r=1}^{11}
(A_r^a w_a)} }
generates the $d=10$ Green's function
$\Phi(x) = (x^\mu x_\mu)^{-4}.$

One can similarly construct massless $d=10$ solutions to higher-spin
equations by using the twistor transform formula
\eqn\tensolthree{ \Phi^{(a_1 ... a_{N})}(x)
= \oint [d\l]_{d=10} \l^{a_1} ...\l^{a_{N}} F(\l,w)|_{w=x\l}}
where $F(\l^a, w_b)$ satisfies the
condition
$F(h\l^a,h w_b) = h^{-N-8} F(\l^a,w_b)$ for $N$ positive.
Since ${\p\over{\p x^\mu}} F(\l,w) = (\l\s_\mu)_a {\p\over{\p w_a}}F(\l,w)$,
one can use $\s^\mu_{ab} \s_{\mu~cd}\l^a\l^c=0$
to show that
$\s^\mu_{bc}{\p\over{\p x^\mu}}
\Phi^{(c a_2 ... a_{N})}(x)=0$.
So \tensolthree\
describes a massless spin $\half$ field when $N=1$,
a self-dual five-form field-strength when $N=2$, etc.
As in the $d=8$ case,
one cannot construct massless solutions when $N$ is negative since
$\s^\mu_{ab} \s_{\mu~cd}{\p\over{\p w_b}}
{\p\over{\p w_d}}$ does not necessarily vanish.

\newsec{Twistor Transform Construction in Higher Dimensions}

In this section, we will generalize the twistor
transform constructions of the previous
sections
to arbitrary even dimension. In dimension $d=2n$,
a pure spinor is defined as a chiral spinor
$\l^a$ for $a=1$ to $2^{n-1}$ which satisfies the additional constraints
\eqn\pureall{\l^a \s^{\mu_1 ... \mu_{n-4}}_{ab} \l^b
=\l^a \s^{\mu_1 ... \mu_{n-8}}_{ab} \l^b=
\l^a \s^{\mu_1 ... \mu_{n-12}}_{ab} \l^b=
   ... =0.}
To covariantize
$u_{jk}$ and $v_j$ of
\genone\ for $j=1$ to $n$,
define the antichiral pure spinor
\eqn\eventwo{ w_{\overline b} = \s^\mu_{a\overline b} x_\mu \l^a}
where $\s^\mu_{a\overline b}$ are the $d=2n$ Pauli matrices and
$\overline b$ denotes $\overline b=\bd$ when $n$ is even and
$\overline b= b$ when $n$ is odd.
Under $d=2n$ conformal transformations,
$(\l^a, w_{\overline b})$ transforms linearly as an $SO(2n,2)$ spinor.

When
\eqn\expand{\l^a = (1,~u_{j_1 j_2},
~-{1\over 8} u_{[j_1 j_2} u_{j_3 j_4]},
~-{1\over {48}} u_{[j_1 j_2} u_{j_3 j_4}u_{j_5 j_6]},~ ...),}
one can choose a representation of the $d=2n$ Pauli matrices
such that
\eqn\evenchoose{ w_{\bar a} = ( v_{j_1},~\half v_{[j_1} u_{j_2 j_3]},~
{1\over 8} v_{[j_1} u_{j_2 j_3} u_{j_4 j_5]},~ ...)}
which satisfies
\eqn\evensat{
\l\s^{\mu_1 ... \mu_{n-3}}w=
\l\s^{\mu_1 ... \mu_{n-5}}w=
\l\s^{\mu_1 ... \mu_{n-7}}w= ... = 0,}
as implied by \eventwo.

To define integration of pure spinors over the coset space
$SO(2n)/U(n)$, a central role
will be played by a
Lorentz-invariant tensor
\eqn\tensordef{ T^{[a_1 ... a_R]  (b_1 ... b_S)}}
which is antisymmetric in its first $R$ indices, symmetric in its
last $S$ indices, and satisfies
$$ T^{[a_1 ... a_R]  (b_1 ... b_S)}
\s^{\mu_1 ... \mu_{n-4}}_{b_1 b_2} =
T^{[a_1 ... a_R]  (b_1 ... b_S)}
\s^{\mu_1 ... \mu_{n-8}}_{b_1 b_2} =
... =0.  $$
When $d=2n$,
$R= 2^{n-1} -1 - n(n-1)/2$ and $S= (n-2)(n-3)/2$.
This tensor can be explicitly constructed by defining
\eqn\Tdef{T^{[a_1 ... a_R] (b_1 ... b_S)} \t_{a_1} ... \t_{a_R}
\tau_{b_1} ... \tau_{b_s} =}
$$ ({\p\over{ \p\tau}}\s^{j_1 ... j_n} {\p\over{ \p\tau}})
({\p\over{ \p\tau}}\s^{j_{n+1} ... j_{2n}} {\p\over{ \p\tau}}) ...
({\p\over{ \p\tau}}\s^{j_{n(n-2)+1} ... j_{n(n-1)}} {\p\over{ \p\tau}}) $$
$$(\tau {\p\over{\p\t}})
(\tau \s_{j_1 j_2}{\p\over{\p\t}})
(\tau \s_{j_3 j_4}{\p\over{\p\t}}) ...
(\tau \s_{j_{n(n-1)-1} j_{n(n-1)}} {\p\over{\p\t}})
~~ (\t)^{2^{n-1}}$$
where $\t_a$ is a fermionic spinor and $\tau_a$ is a
bosonic pure spinor.\foot{It is interesting to note
that \Tdef\ is the state with maximum number of $\tau$'s
in the cohomology of the nilpotent operator $Q=\tau_a {\p\over{\p\t_a}}$.
When $n=5$, $Q$ is the zero-momentum contribution to the BRST operator
for the $d=10$ superparticle.\ref\superpart{N. Berkovits,
``Covariant quantization of the superparticle using
pure spinors,'' JHEP 09 (2001) 016, hep-th/0105050.}}
Note that there are $(2n-2)$ ${\p\over{\p\tau}}$'s,
${{n^2-n+2}\over 2}$ $\tau$'s,
${{n^2-n+2}\over 2}$
${\p\over{\p\t}}$'s, and $2^{n-1}$ $\t$'s on the right-hand side of
\Tdef, which agrees with the powers of $\tau$ and $\t$
on the left-hand side of \Tdef.
When $n=4$, $T^{ab}\tau_b$ is proportional to
$\tau^a$,
and when $n=5$, $T^{[a_1 ... a_5](b_1 b_2 b_3)}\tau_{b_1}\tau_{b_2}
\tau_{b_3}$ is proportional to
$(\s^\mu\tau)^{a_1} (\s^\nu\tau)^{a_2} (\s^\rho\tau)^{a_3}
(\s_{\mu\nu\rho})^{a_4 a_5}.$

To define integration over $\l^a$, note
that the measure factor
\eqn\evenmeasure{ [d\l]_{d=2n}\equiv
(C_f \l^f)^{-S}
\e_{a_1 ... a_{n(n-1)/2} b c_1 ... c_R} d\l^{a_1}
\wedge ... \wedge d\l^{a_{n(n-1)/2}} ~\l^b
T^{[c_1 ... c_R](e_1 ... e_S)}
C_{e_1} ... C_{e_S} }
is independent of the choice of $C_b$ and is therefore Lorentz-invariant.
As in the $d=10$ case described in the previous section, the easiest
way to prove independence of \evenmeasure\ on $C_b$
is to use
the invariance under the $U(1)\times SU(n)$ subgroup
which preserves the pure spinor $\l^a$ up to a phase. Under $U(1)\times
SU(n)
$,
the components of an $SO(2n)$
chiral spinor transform with $U(1)$ charges
$(n/2, (n-4)/2, (n-8)/2,  ...)$,
and $\l^a$ transforms with $U(1)$ charge $n/2$.
Furthermore,
since $\l^a \s^{\mu_1 ... \mu_{n-4}} d\l=
\l^a \s^{\mu_1 ... \mu_{n-8} } d\l= ... = 0$,
$d\l^a$ carries either $U(1)$ charge $n/2$ or $(n-4)/2$. Therefore,
$d\l^{[a_1}
\wedge ... \wedge d\l^{a_{n(n-1)/2}} ~\l^{b]}$ carries $U(1)$ charge
$n(n-2)(n-3)/4$,
which implies by $U(1)$ conservation
that only the component of $C_b$
with $U(1)$ charge $-n/2$ contributes to \evenmeasure. Finally,
it is easy to see that
\evenmeasure\ is invariant under scale transformations of this $-n/2$
component because there is an equal number of $C_b$'s in the numerator
and denominator of \evenmeasure.

Using the measure factor of \evenmeasure,
the solution of \gensol\ can be written
in Lorentz-covariant form as the twistor transform formula
\eqn\evensoltwo{ \Phi(x) = \oint [d\l]_{d=2n}
F(\l,w)|_{w=x \l}}
where $F(h\l^a,h w_{\overline b}) = h^{2-2n}
F(\l^a,w_{\overline b})$ so that the integral
over the projective spinor is well-defined.
For example, choosing
\eqn\evenextwo{ F(\l,w) =
{\e_{\bar b_1 ... \bar b_M \bar c_1 ...\bar c_R}
A_1^{\bar b_1} ... A_M^{\bar b_M}
T^{[\bar c_1 ... \bar c_R] (\bar e_1 ... \bar e_S)}
w_{\bar e_1} ... w_{\bar e_S}\over \prod_{j=1}^M
(A_j^{\bar a} w_{\bar a})}}
generates the $d=2n$ Green's function
$\Phi(x) =
(x^\mu x_\mu)^{1-n}$
where $M=(n^2-n+2)/2$.

One can similarly construct massless $d=2n$ solutions to higher-spin
equations by using the twistor transform formula
\eqn\evensolthree{ \Phi^{(a_1 ... a_{N})}(x)
= \oint [d\l]_{d=2n} \l^{a_1} ...\l^{a_{N}} F(\l,w)|_{w=x\l}}
where $F(\l^a, w_{\bar b})$ satisfies the
condition
$F(h\l^a,h w_{\bar b}) = h^{-N +2-2n} F(\l^a,w_{\bar b})$ for $N$ positive.
Since ${\p\over{\p x^\mu}} F(\l,w) = (\l\s_\mu)_{\bar a}
{\p\over{\p w_{\bar a}}}F(\l,w)$,
one can use $\s^\mu_{a{\bar b}} \s_{\mu~c{\bar d}}\l^a\l^c=0$
to show that
$\s^\mu_{b{\bar c}}{\p\over{\p x^\mu}}
\Phi^{(b a_2 ... a_{N})}(x)=0$.
So \evensolthree\
describes a massless spin $\half$ field when $N=1$,
a self-dual $n$-form field-strength when $N=2$, etc.
Unlike the $d=4$ and $d=6$ cases,
one cannot construct massless solutions when $N$ is negative since
$\s^\mu_{a{\bar b}} \s_{\mu~c{\bar d}}{\p\over{\p w_{\bar b}}}
{\p\over{\p w_{\bar d}}}$ does not necessarily vanish.

\bigskip
\bigskip
\bigskip
\noindent {\bf Acknowledgments}
We are grateful to Paul Howe,
Nikita Nekrasov and Edward Witten for useful conversations
and the Simons Workshop in Mathematics
and Physics at SUNY at Stony Brook
for their hospitality. 
We are grateful also to L. P. Hughston for helpful comments on an
earlier version of this paper after it appeared on the hep-th archive.
NB was partially supported by CNPq grant 300256/94-9,
Pronex 66.2002/1998-9 and FAPESP grant 99/12763-0.
SCh was supported in part by DOE grant DE-FG02-90ER40542.

\bigskip
\bigskip
\bigskip
\noindent {\bf Appendix}

Here we collect some facts about pure spinors, elucidating their relation to
a few
descriptions of conventional twistors. In particular, we discuss the
relation of pure
spinors to complex structures on ${\bf R}^{2n},$ as well as to isotropic
complex
Grassmanians.

\subsec{Complex Structures on ${\bf R}^{2n}$}
Consider all complex structures on
${\bf R}^{2n}$ that are compatible with the flat metric. Since these are
produced by all
orthonormal changes of coordinates modulo the complex changes of
coordinates,
the moduli
space of all complex structures is $SO(2n)/U(n).$

Let us write this in more detail. Identifying ${\bf R}^{2n}$ with ${\bf C}
^n,$
with
$z\in{\bf C}^n$ given by coordinates $z_i$ so that $z=(z_1,\ldots,
z_n),$ the
metric is
given by
$$ds^2=(z,\bar{z})G \pmatrix{ z \cr \bar{z}},\ G=\pmatrix{ 0&1_n \cr
1_n&0}.$$
If $M\in SO(2n)$,
then $M$ satisfies $M M^\dagger=1$ and
$\overline{M}=G M G$
in the $(z,\bar{z})$ basis.
That is, $M$ can be written in a block form
$$ M=\pmatrix{ T&U \cr \bar{U}&\bar{T}},$$
with
\eqn\norm{T T^\dagger+U U^\dagger=1,}
\eqn\other{T U^t+U T^t  =  0.}
Such a matrix $M$ defines complex
coordinates $v=T z+U \bar{z}.$ Unitary transformations
$$\pmatrix{\Lambda&0 \cr 0&\bar{\Lambda}},\
\Lambda\Lambda^\dagger=1, $$
respect the complex structure, $v\sim\Lambda v,$ and
generically can be used to put $T=1_n$. This can
be compared to formula \genone\ for $v_j$ in the text.

\subsec{Isotropic Grassmanian}
Dropping the normalization condition \norm\ and considering now $\Lambda\in
GL(n,{\bf C})$, we observe that there is a unique solution to \norm\ on the
$GL(n,{\bf C})$
orbit. Thus $SO(2n)/U(n)$ can be thought of as the space of pairs
$(T,U)\sim(\Lambda T,
\Lambda U)$ with
$$\eqalign{
\det(T T^\dagger+U U^\dagger)\quad \neq \quad 0 \cr
T U^t+U T^t \quad = \quad 0.}
$$
The latter is the space of isotropic n-planes in ${\bf C}^{2n}.$ That
is, the
$2n\times n$
matrix $\pmatrix{ T \cr U }$ defines ${\bf C}^n\subset{\bf C}^{2n}$ that
is isotropic since
$$
(T,U) G \pmatrix{ T^t \cr U^t }=0.
$$
Thus
$$ SO(2n)/U(n)=G^0_n({\bf C}^{2n}),$$
the isotropic Grassmanian. This space is of real dimension $n(n-1)$.

Given
$(T,U )^t$, we
consider a coordinate patch with $\det U\neq0$. Then $(T,U)^t\sim(U^{-1}T,
1)^t$ and the
elements of the antisymmetric
matrix $U^{-1}T$ provide coordinates in the patch. A
different choice of basis amounts to a permutation of rows in $(T,U)^t.$
There
are exactly
$2^{n-1}$ such permutations respecting the orientation of the space.
Thus the
isotropic
Grassmanian $G^0_n({\bf C}^{2n})$ is covered by a minimum of $2^{n-1}$
coordinate charts, in other words
its Lusternik-Schnirelmann category is $2^{n-1}.$

\subsec{Pure Spinors} For a pure spinor $\lambda^\alpha$ the only
nonvanishing
form is
of degree $n$. Moreover, it is simple, i.e. its coefficients satisfy
$\lambda^\alpha\gamma^{i_1\ldots
i_n}_{\alpha\beta}\lambda^\beta=a_1^{[i_1}a_2^{i_2}\ldots a_n^{i_n]}$ for
some complex
linearly independent vectors $a_1, a_2,\ldots a_n.$ Thus each pure spinor
defines a
complex n-plane in ${\bf C}^{2n}.$ Moreover, this plane is isotropic since
$$
g_{i_1 j_1}(\lambda^\alpha\gamma^{i_1\ldots
i_n}_{\alpha\beta}\lambda^\beta)(\lambda^\gamma\gamma^{j_1\ldots
j_n}_{\gamma\delta}\lambda^\delta)=0.
$$
This correspondence is known as the Cartan map. It is one-to-one for
projective pure
spinors. Thus the space of all projective pure spinors is the isotropic
Grassmanian
$G^0_n({\bf C}^{2n}).$ We conclude that projective pure spinors
parameterize complex
structures on ${\bf R}^{2n}.$

\bigskip
\bigskip
\bigskip

\listrefs
\end